
\input phyzzx.tex
\input tables.tex

\def\MPL #1 #2 #3 {{\sl Mod.~Phys.~Lett.}~{\bf#1} (#2), #3}
\def\NPB #1 #2 #3 {{\sl Nucl.~Phys.}~{\bf B#1} (#2), #3}
\def\PLB #1 #2 #3 {{\sl Phys.~Lett.}~{\bf#1} (#2), #3}
\def\PR #1 #2 #3 {{\sl Phys.~Rep.}~{\bf#1} (#2), #3}
\def\PRD #1 #2 #3 {{\sl Phys.~Rev.}~{\bf D#1} (#2), #3}
\def\PRL #1 #2 #3 {{\sl Phys.~Rev.~Lett.}~{\bf#1} (#2), #3}
\def\RMP #1 #2 #3 {{\sl Rev.~Mod.~Phys.}~{\bf#1} (#2), #3}
\def\ZP #1 #2 #3 {{\sl Z.~Phys.}~{\bf#1} (#2), #3}
\def\IJMP #1 #2 #3 {{\sl Int.~J.~Mod.~Phys.}~{\bf#1} (#2), #3}
\def\CPC #1 #2 #3 {{\sl Comp.~Phys.~Comm.}~{\bf#1} (#2), #3}
\def\CP #1 #2 #3 {{\sl Comp.~Phys.}~{\bf#1} (#2), #3}

\def\frac#1#2{{#1 \over #2}}

\def\mw{{m_{\scriptscriptstyle W}}}
\def\M{{\cal M}}
\def\s{\sigma}
\def\th{\theta}
\def\la{\lambda}
\def\eps{\epsilon}
\def\LL{{\scriptscriptstyle L}}
\def\RR{{\scriptscriptstyle R}}
\def\dslash{\not{\hbox{\kern-2pt $\partial$}}}
\def\Dslash{\not{\hbox{\kern-4pt $D$}}}
\def\Qslash{\not{\hbox{\kern-4pt $Q$}}}
\def\pslash{\not{\hbox{\kern-2.3pt $p$}}}
\def\kslash{\not{\hbox{\kern-2.3pt $k$}}}
\def\qslash{\not{\hbox{\kern-2.3pt $q$}}}
\def\Vslash{\not{\hbox{\kern-2pt $V$}}}
\def\trace#1{{\rm tr} \left\{#1\right\}}
\def\bp{{\bf p}}
\def\bs{{\bf\s}}

\PHYSREV
\contentson
\tableofplateson

\date={September 1992}
\pubnum={92/256-T}
\titlepage
\title{The Vector Equivalence Technique}
\author{Eran Yehudai\footnote*{Internet: eran@fnth06.fnal.gov}}
\FNAL
\abstract
We present the Vector Equivalence technique. This technique allows a
simple and systematic calculating of Feynman diagrams involving
massive fermions at the matrix element level. As its name suggests,
the technique allows two Lorentz four-vectors to serve as an
equivalent of two external fermions. In further calculations, traces
involving these vectors replace the matrix element with the external
fermions. The technique can be conveniently used for both symbolic and
numeric calculations.
\submit{Physical Review D}
\endpage

\chapter{Introduction}

Calculations of Feynman diagrams with external fermions occur
frequently in particle phenomenology. The traditional calculation
technique calls for squaring the amplitude while summing over
polarizations. This method has the advantage that the final expression
involves only dot-products of Lorentz vectors and, possibly,
contraction with the Levi-Civita tensor. A major disadvantage of this
method is that the number of terms in the result grows as the square
of the number of terms in the amplitude. Both in tree-level and in
higher order calculations, this can be a severe shortcoming.

Several authors have proposed methods for calculating the matrix
element without squaring
\REF\Hagiwara{K. Hagiwara and D. Zeppenfeld, \NPB 274 1986 1.}
\REF\SpinTech{H. W. Fearing and R. B. Silbar, \PRD 6 1972 471;
\nextline R. Kleiss, W. J. Stirling, \NPB 262 235 1985.}
\refmark{\Hagiwara, \SpinTech}. We
\REF\Form{J.~A.~M.~Vermaseren, Report No.~KEK-Th-326 (1992).}
propose yet another such method. Its main advantage is that it gives,
much like the traditional method, a relatively simple symbolic
expression of $\M$ even when massive fermions are involved. Unlike
other methods, one can perform calculations with free Lorentz indices.
A similar technique has been implemented using the symbolic language
Form.\Form

Generally speaking, the method entails substituting for each pair of
external fermions, two complex Lorentz vectors, corresponding to the
vector and pseudo-vector currents. Any amplitude
involving the two fermions can then be rewritten as a trace involving
the various four-vectors (and free Lorentz indices) in the problem and
these two new four-vectors.

\REF\HIP{A.~Hsieh and E.~Yehudai, \CP 6 1992 253;\nextline
E.~Yehudai, Report No.~Fermilab-Pub-92/22-T (1992)}
\REF\FeynCalc{R.~Mertig, M.~B\"ohm and A.~Denner, \CPC 64 1991 345.}
\REF\Dissertation{E.~Yehudai, Report No.~SLAC-Report-383, 1991
(Ph.D. dissertation).}
The Vector Equivalence technique was first described, and used
extensively, in ref.~\Dissertation. The technique can easily be
combined with computerized packages for symbolic manipulation of the
Dirac algebra\refmark{\HIP, \FeynCalc}. The Vector Equivalence
technique is already implemented in the package described in
ref.~\HIP, and can easily be added to other packages.

This paper proceeds as follows. In the next section we derive the
Vector Equivalence technique. We describe how to use the two currents
to rewrite arbitrary amplitudes, and quote some useful identities. In
sec.~3 we give an example for the use of the Vector Equivalence
technique. We use it to calculate the helicity amplitude for the
process $e^+e^-\rightarrow W^+W^-$ in a model which includes excited
neutrinos\Ref\ExcitedFermions{K.~Hagiwara, S.~Komamiya and
D.~Zeppenfeld, \ZP C29 1985 115.}.
The excited neutrinos couple to the
electrons via a magnetic dipole transition. The methods described in
refs.~1-2 cannot be simply used to derive this result. In sec.~4 we
present our conclusions. In order to calculate the actual vector
currents, one has to resort to an explicit representation of the
spinors. In the appendix we describe, for completeness, one such
representation, closely following ref.~1.

\chapter{The Vector Equivalence Technique}

In calculating a Feynman diagram with external fermions, one encounters
objects of the form
$$
\M = \bar u(p, s) \Gamma u(p',s'), \eqn\defGamm
$$
where $p$ and $p'$ are the momenta of the external fermions, $s$
and $s'$ are their helicities, and $\Gamma$ is an arbitrary string of
Dirac gamma matrices. For simplicity, we are only referring to
fermions (as opposed to anti-fermions) in this derivation. For the
purpose of this discussion, an anti-fermion with mass $m$ behaves
exactly like a fermion with mass $-m$. Additionally, we
suppress the reference to $s$ and $s'$ in the derivation.

The traditional method calls for squaring $\M$ while summing over
fermion helicities:
$$
\sum_{s,s'} |\M|^2 = \sum_{s,s'} \trace{\bar u(p,s) \Gamma
u(p',s')\bar u(p',s') \bar\Gamma u(p,s)} =
\trace{(\pslash+m)\Gamma(\pslash'+m')\bar\Gamma}, \eqn\trad
$$
where $\gamma^0\bar\Gamma = {\Gamma^R}^\ast\gamma^0$ and $m$ and $m'$
are the masses of $p$ and $p'$ respectively.
This method is advantageous in that the final result is expressed in
terms of easy-to-calculate Lorentz invariants. However, it becomes
cumbersome as the number of terms in $\M$ increases.

We start by rewriting
$$
\M=\bar u(p) \Gamma u(p') = \trace{\Gamma u(p') \bar u(p)}.
$$
Next, express $u(p') \bar u(p)$ in terms of an orthogonal basis
$\big\{\Gamma^{(i)}\big\}$ of the four dimensional Dirac space.
This basis obeys the orthonormality relation
$$
\trace{\Gamma^{(i)}{\Gamma^{(j)}}^R} = \delta^{ij}.
$$
In terms of such a basis, one can write
$$
u(p')\bar u(p) = \sum_{(i)} V^{(i)} \Gamma^{(i)}.
$$
The coefficients $V^{(i)}$ can be calculated using a projection:
$$
V^{(i)} = \trace{u(p')\bar u(p) {\Gamma^{(i)}}^R} = \bar
u(p) {\Gamma^{(i)}}^R u(p'). \eqn\VviaTrace
$$
Given $V^{(i)}$, $\M$ may be written as
$$
\M = \bar u(p) \Gamma u(p') = \sum_{(i)} V^{(i)}
\trace{\Gamma\Gamma^{(i)}}. \eqn\MviaV
$$
This equation can be simplified if we consider the fact
that $p$ and $p'$ represent on-shell fermions obeying the Dirac
equation\Ref\Vermaseren{I would like to thank J. Vermaseren for his
help in pointing out the usefulness of these identities.}:
$$
\bar u(p)(\pslash-m) = (\pslash'-m') u(p') = 0.
$$
For any $\Gamma^{(i)}$ we can write
$$\eqalign{
0 &= \bar u(p) (\pslash-m) {\Gamma^{(i)}}^R u(p')
= \sum_{(j)} V^{(j)} \trace{(\pslash-m) {\Gamma^{(i)}}^R
\Gamma^{(j)}}, \cr
0 &= \bar u(p) {\Gamma^{(i)}}^R (\pslash'-m') u(p')
= \sum_{(j)} V^{(j)} \trace{{\Gamma^{(i)}}^R (\pslash'-m')
\Gamma^{(j)}},
}$$
or
$$\eqalign{
m V^{(i)} &= \sum_{(j)} V^{(j)} \trace{\pslash{\Gamma^{(i)}}^R
\Gamma^{(j)}}, \cr
m' V^{(i)} &= \sum_{(j)} V^{(j)} \trace{{\Gamma^{(i)}}^R \pslash'
\Gamma^{(j)}}.\cr} \eqn\eqnmotion
$$

Let us now consider a particular choice for the basis
$\left\{\Gamma^{(i)}\right\}$, namely
$$
\left\{\Gamma^{(i)}\right\} = \left\{\frac{1}{2}, \frac{\gamma^\mu}{2},
\frac{\gamma^{\mu\nu}}{2\sqrt{2}}, \frac{\gamma^5\gamma^\mu}{2},
\frac{\gamma^5}{2}
\right\},
$$
where $\gamma^{\mu\nu} =
(\gamma^\mu\gamma^\nu-\gamma^\nu\gamma^\mu)/2$. The corresponding
$\left\{V^{(i)}\right\}$ are
$$
\left\{V^{(i)}\right\} = \left\{U, V^\mu, W^{\mu\nu}, V^\mu_5,
U_5\right\}.
$$
Equation~\MviaV\ then takes the form
$$
\bar u(p)\Gamma u(p') = \frac{1}{2} \trace{\Gamma (U + \Vslash +
\frac{1}{\sqrt{2}}W_{\mu\nu} \gamma^{\mu\nu} + \gamma^5 \Vslash_5 +
U_5 \gamma^5)}. \eqn\MviaUV
$$

The string $\Gamma$ of equation~\defGamm\ can always be written
as a sum $\Gamma=\Gamma_{\rm odd}+\Gamma_{\rm even}$ where
$\Gamma_{\rm odd}$ and $\Gamma_{\rm even}$ contain an even and an odd
number of gamma matrices respectively. Equation~\MviaUV\ can be
broken into
$$
\bar u(p) \Gamma_{\rm odd} u(p') = \frac{1}{2}
\trace{\Gamma_{\rm odd} (\Vslash+\gamma^5\Vslash_5)}, \eqn\OddFinal
$$
$$
\bar u(p) \Gamma_{\rm even} u(p') = \frac{1}{2}
\trace{\Gamma_{\rm even} (U+U_5 \gamma^5 + \frac{1}
{\sqrt{2}} W_{\mu\nu} \gamma^{\mu\nu})}. \eqn\GammEve
$$

If both fermions are massless, the string $\Gamma$ of
equation~\defGamm\ has to contain an odd number of gamma
matrices, and we therefore have $U=U_5=W^{\mu\nu}=0$.
Let us assume then that $m\ne 0$. Substituting $\Gamma^{(i)} =
\gamma^{\mu\nu}/2\sqrt{2}$ into equation~\eqnmotion\ gives
$$\eqalign{
W^{\mu\nu} &= \frac{1}{4\sqrt{2}m}\trace{\pslash \gamma^{\nu\mu}
(\Vslash + \gamma^5\Vslash_5)} = \frac{1}{\sqrt{2}m} (p^\nu V^\mu -
p^\mu V^\nu + i\epsilon^{\mu\nu p V_5}), \cr
&= \frac{1}{4\sqrt{2}m'}\trace{\gamma^{\nu\mu} \pslash'
(\Vslash + \gamma^5 \Vslash_5)} = \frac{1}{\sqrt{2}m'} ({p'}^\mu V^\nu
- {p'}^\nu V^\mu + i\epsilon^{\mu\nu p' V_5}),} \eqn\WviaV
$$
where $\epsilon^{\mu\nu V_5 p}$ is a shorthand for
$\epsilon^{\mu\nu\alpha\beta} {V_5}_\alpha p_\beta$.
Similarly, substituting $U$ and $U_5$ for $\Gamma^{(i)}$ gives
$$\eqalign{
U &= \frac{1}{4m} \trace{\pslash (\Vslash+\gamma^5 \Vslash_5)} =
\frac{V\cdot p}{m} = \frac{V\cdot p'}{m'} \cr
U_5 &= \frac{1}{4m} \trace{\pslash \gamma^5
(\Vslash+\gamma^5 \Vslash_5)} = \frac{V_5\cdot p}{m} = -\frac{V_5\cdot
p'}{m'}.} \eqn\UviaV
$$
Using equation~\WviaV\ and \UviaV, equation~\GammEve\ takes the form:
$$
\bar u(p) \Gamma_{\rm even} u(p') =
\frac{1}{2m}\trace{\Gamma_{\rm even} (\Vslash+\gamma^5\Vslash_5)
\pslash} \eqn\EvenFinal
$$
or
$$
\bar u(p) \Gamma_{\rm even} u(p') =
\frac{1}{2m'}\trace{\Gamma_{\rm even} \pslash'
(\Vslash+\gamma^5 \Vslash_5)} \eqn\EvenFinalII
$$

Equations~\OddFinal\ and \EvenFinal\ (or
\EvenFinalII) are all one needs
to calculate the generic matrix element $\M$ of
equation~\defGamm\ in terms of the two four-vectors $V^\mu$ and
$V_5^\mu$. $V^\mu$ and $V_5^\mu$ depend on the four-vectors $p$ and
$p'$ and the helicities $s$ and $s'$. Since we do not implicitly sum
over fermion helicity, this summation has to be carried out
explicitly.

When the fermions are involved in chiral interactions such as
electro-weak interactions, it is often more convenient to use a chiral
basis for the Dirac space:
$$
\left\{\Gamma^{(i)}\right\} = \left\{\frac{P_{\LL,\RR}}{\sqrt{2}},
\frac{\gamma^\mu P_{\LL,\RR}}{\sqrt{2}},
\frac{\gamma^{\mu\nu}}{2\sqrt{2}}\right\},\eqn\ChiralBase
$$
where $P_\LL = (1-\gamma^5)/2$ and $P_\RR = (1+\gamma^5)/2$.
The corresponding four-vectors $V_\LL$ and $V_\RR$ are related
to $V$ and $V_5$ via
$$
\eqalign{
&V_\RR = \frac{1}{\sqrt{2}} (V-V_5), \qquad
V_\LL = \frac{1}{\sqrt{2}} (V+V_5), \cr
&V=\frac{V_\LL+V_\RR}{\sqrt{2}}, \qquad
V_5=\frac{V_\LL-V_\RR}{\sqrt{2}}.}\eqn\VLeftRight
$$
In terms of $V_\LL$ and $V_\RR$, equations~\OddFinal,
\EvenFinal\ and \EvenFinalII\ take the form
$$\eqalign{
\bar u(p) \Gamma_{\rm odd} u(p') &=
\frac{1}{\sqrt{2}}\trace{\Gamma_{\rm odd} (\Vslash_\RR P_\RR +
\Vslash_\LL P_\LL)}, \cr
\bar u(p) \Gamma_{\rm even} u(p') &=
\frac{1}{\sqrt{2}m}\trace{\Gamma_{\rm even} (\Vslash_\RR P_\RR +
\Vslash_\LL P_\LL) \pslash}, \cr
\bar u(p) \Gamma_{\rm even} u(p') &=
\frac{1}{\sqrt{2}m'}\trace{\Gamma_{\rm even} \pslash' (\Vslash_\RR
P_\RR + \Vslash_\LL P_\LL)}.} \eqn\EvenOddChiral
$$

The entire derivation thus far did not depend on any particular
representation of gamma matrices or spinors. In order to express the
four-vectors $V$ and $V_5$ (or $V_\LL$ and $V_\RR$) in terms of the
fermion momenta $p$ and $p'$, one needs to settle on a particular
representation. For the specific cases of $V$, $V_5$ and $V_\la$,
equation~\VviaTrace\ gives
$$\eqalign{
V^\mu &= \frac{1}{2} \bar u(p,s) \gamma^\mu u(p',s'),
 \cr
V^\mu_5 &= \frac{1}{2} \bar u(p,s) \gamma^\mu \gamma^5 u(p',s'),
 \cr
V^\mu_\lambda &= \frac{1}{\sqrt{2}} \bar u(p,s) \gamma^\mu P_\lambda
u(p',s').} \eqn\VviaG
$$
coupled with a specific representation, these equations form a
prescription for calculating $V$, $V_5$ and $V_\la$.
Calculating $V_\la$ is particularly convenient if one chooses a chiral
representation for the spinors, such as the one described in the
appendix.

Finally, we would like to collect several identities involving the
$V$'s which can be used in simplifying and verifying
calculations. From equation~\WviaV\ one gets
$$
i\epsilon^{\mu\nu\alpha V_5}
\left(\frac{p}{m}-\frac{p'}{m'}\right)_\alpha = \left(\frac{p}{m} +
\frac{p'}{m'}\right)^\mu V^\nu - \left(\frac{p}{m} +
\frac{p'}{m'}\right)^\nu V^\mu.
$$
{}From equations~\UviaV\ follows:
$$
V\cdot\left(\frac{p}{m}-\frac{p'}{m'}\right) = 0, \qquad
V_5\cdot\left(\frac{p}{m}+\frac{p'}{m'}\right) = 0.
$$
When squaring an expression involving the $V$'s, one can make use of
equation~\trad\ to arrive at the following identities:
$$\eqalign{
&\sum_{ss'}V_\mu V^\ast_\nu = p_\mu p'_\nu + p_\nu p'_\mu - ((p\cdot
p') -mm') g_{\mu\nu} \cr
&\sum_{ss'}V^5_\mu {V^5}^\ast_\nu = p_\mu p'_\nu + p_\nu p'_\mu -
((p\cdot p') +mm') g_{\mu\nu} \cr
&\sum_{ss'}i\epsilon^{\mu\nu\alpha\beta} V^5_\alpha V^\ast_\beta =
2(p^\mu {p'}^\nu -p^\nu {p'}^\mu).
}$$

\chapter{Example: $e^+e^-\rightarrow W^+W^-$ with Excited Fermions}

In this section we present one calculation carried out with the Vector
Equivalence technique. We chose to calculate one helicity amplitude
for the process $e^+e^-\rightarrow W^+W^-$ in a model which extends
the Standard Model by including an excited neutrino. The excited
neutrino is massive, and couples to the $W$ and electron via a
magnetic transition\refmark\ExcitedFermions. The relevant effective
Lagrangian is:

$$
{\cal L_{\rm eff}} = \frac{e}{\Lambda} \bar \nu^\ast \sigma^{\mu\nu}
(c-d\gamma_5) e \partial_\mu W_\nu + {\rm h.c.},\eqn\Leff
$$
where $\Lambda$ is the compositeness scale.
While the electron mass can normally be neglected in
high-energy collisions, we keep it finite to illustrate the treatment
of massive external fermions.

The matrix element for the process is given by
$$\eqalign{
\M^{\la^-\la^+}_{\s^-\s^+} &= \eps^{\mu\ast}(p_3,\la^-)
\eps^{\nu\ast}(p_4,\la^+) \times \cr
&
\left(\M_{\s^-\s^+}^{\mu\nu}(\nu) +
\M_{\s^-\s^+}^{\mu\nu}(\gamma) + \M_{\s^-\s^+}^{\mu\nu}(Z) +
\M_{\s^-\s^+}^{\mu\nu}(H) + \M_{\s^-\s^+}^{\mu\nu}(\nu^\ast)\right),
}\eqn\Mll
$$
where $\s^-$, $\s^+$, $\la^-$ and $\la^+$ are the helicities of the
electron, positron, $W^-$ and $W^+$ respectively,
$$\eqalign{
\M^{\mu\nu}_{\s^+\s^-}(\nu) &= \bar v_{\s^+}(p_2)
\left(\frac{ig}{\sqrt{2}}\gamma_\nu P_\LL \right) \frac{i(\pslash_1 -
\pslash_3+m_e)}{t-m_e^2} \left(\frac{ig}{\sqrt{2}}\gamma_\mu
P_\LL\right) u_{\s^-}(p_1) \crr\cr
\M^{\mu\nu}_{\s^+\s^-}(\gamma) &= \bar v_{\s^+}(p_2)
(-ie\gamma_\alpha) u_{\s^-}(p_1)
\left(\frac{-ig^{\alpha\tau}}{s}\right) (i \Gamma^\gamma_{\tau\mu\nu})
\crr\cr
\M^{\mu\nu}_{\s^+\s^-}(Z) &= \bar v_{\s^+}(p_2)
(ie\gamma_\alpha(g_0 + g_\LL P_\LL)) u_{\s^-}(p_1)
\left(\frac{-ig^{\alpha\tau}}{s-m_Z^2}\right) (i \Gamma^Z_{\tau\mu\nu})
\crr\cr
\M^{\mu\nu}_{\s^+\s^-}(H) &= \bar v_{\s^+}(p_2)
\left(\frac{-igm_e}{2m_W}\right) u_{\s^-}(p_1)
\frac{i}{s-m_H^2} (im_Wgg^{\mu\nu}) \crr\cr
\M^{\mu\nu}_{\s^+\s^-}(\nu^\ast) &= \bar v_{\s^+}(p_2)
\s^{\nu\beta} \frac{ie}{\Lambda} (c-d\gamma_5)\left(\frac{i(\pslash_3-
\pslash_1+m_{\nu^\ast})}{t-m_{\nu^\ast}^2}\right)\times \cr
& \qquad\qquad\qquad \s^{\mu\alpha}
\frac{ie}{\Lambda} (c-d\gamma_5) u_{\s^-}(p_1) p_3^\alpha p_4^\beta.
} \eqn\MMs $$
Here $\eps^{\mu\ast}(p_3,\la^-)$ and $\eps^{\nu\ast}(p_4\la^+)$ are the
polarization vectors of the $W^-$ and $W^+$ respectively, $s =
(p_1+p_2)^2$, $t = (p_1-p_3)^2$,
$$
g_0 = \frac{\sin\theta_W}{\cos\theta_W},\qquad\qquad
g_\LL = -\frac{1}{2\sin\theta_W\cos\theta_W},\eqn\gZeroLeft
$$
and
$$
\Gamma_V^{\tau\mu\nu} = g_V \left((p_3-p_4)^\tau g^{\mu\nu}
+ 2p_4^\mu g^{\nu\tau} - 2p_3^\nu g^{\mu\tau}\right),\eqn\GammaV
$$
with $g_\gamma = e$ and $g_Z = e \cot\theta_W$.

In the $e^+e^-$ center-of-mass frame, the momenta in the process take
the following values:
$$\eqalign{
p_1 = \frac{\sqrt{s}}{2} (0, 0, \beta_e, 1) &\qquad
p_3 = \frac{\sqrt{s}}{2} (\beta_W \sin\th,0,\beta_W \cos\th, 1)
\cr
p_2 = \frac{\sqrt{s}}{2} (0, 0, -\beta_e, 1) &\qquad
p_4 = \frac{\sqrt{s}}{2} (-\beta_W \sin\th,0,-\beta_W \cos\th, 1),
}\eqn\ExplicitPs
$$
where $\beta_e=\sqrt{1-4 m_e^2/s}$ and $\beta_W=\sqrt{1-4 \mw^2/s}$.
The W polarization vectors are:
$$\eqalign{
\eps^\ast(p_3,\pm) = \frac{1}{\sqrt{2}} (\cos\th,\pm i, {-}\sin\th, 0)
&\qquad
\eps^\ast(p_4,\pm) = \frac{1}{\sqrt{2}} (\cos\th, \mp i, {-}\sin\th,
0) \cr
\eps^\ast(p_3, 0) = \frac{\sqrt{s}}{2\mw} (\sin\th, 0, \cos\th, \beta_W)
&\qquad
\eps^\ast(p_4, 0) = \frac{\sqrt{s}}{2\mw} ({-}\sin\th, 0, {-}\cos\th,
\beta_W).
}\eqn\ExplicitEpss
$$
The vectors $V^{\s^+\s^-}$ and $V^{\s^+\s^-}_5$ are given in
\TABLE\?{The vectors $V^{\s^+\s^-}$ and $V^{\s^+\s^-}_5$} table \?.
\midinsert
\vbox{
{\noexpand\centerline{\Tenpoint Table \?.\enspace
The vectors $V^{\s^+\s^-}$ and $V^{\s^+\s^-}_5$}
\vskip\the\captionskip}
\thicksize=0.6pt
\tablewidth=\hsize
\begintable
$(\s^+\s^-)$ & $V^{\s^+\s^-}$ & $V_5^{\s^+\s^-}$ \cr
$(++)$ & $(0,0,0,{-}m_e/\sqrt{s})$ & $(0,0,0,{-}m_e/\sqrt{s})$ \nr
$(+-)$ & $(\beta_e\sqrt{s}/2, {-}i\beta_e\sqrt{s}/2, 0, 0)$ & $
(\beta_e\sqrt{s}/2, {-}i\beta_e\sqrt{s}/2, 0, 0)$ \nr
$(-+)$ & $({-}\beta_e\sqrt{s}/2, {-}i\beta_e\sqrt{s}/2, 0, 0)$ &
$({-}\beta_e\sqrt{s}/2, {-}i\beta_e\sqrt{s}/2, 0, 0)$ \nr
$(--)$ & $(0,0,m_e/\sqrt{s}, 0)$ & $(0,0,0,{-}m_e/\sqrt{s})$
\endtable}
\endinsert

Applying eqns.~\OddFinal\ and \EvenFinal\ to
eqn.~\MMs\ gives (dropping the $\s^+\s^-$ for convinience):
$$\eqalign{
\M^{\mu\nu}(\nu) &= -i\frac{g^2}{4(t-m_e^2)} \trace{\gamma^\nu
P_\LL (\pslash_1-\pslash_3) \gamma^\mu (\Vslash + \Vslash^5 \gamma_5)}
\cr
&= i\frac{g^2}{\sqrt{2}(t-m_e^2)} \big(-i \eps^{\mu\nu\alpha\beta}
(p_1-p_3)_\alpha {V_\LL}_\beta + (p_1-p_3)\cdot {V_\LL} g^{\mu\nu}
\cr
&- (p_1-p_3)^\mu {V_\LL}^\nu - (p_1-p_3)^\nu V_{\LL}^\mu\big)
\crr\cr
\M^{\mu\nu}(\gamma) &= -i\frac{2e}{s} \Gamma^\gamma_{\mu\nu\tau}
V^\tau \crr\cr
\M^{\mu\nu}(Z) &= -i\frac{2e}{s} \Gamma^Z_{\mu\nu\tau}
(g_0 V^\tau + \sqrt{2} g_\LL V_\RR^\tau)\crr\cr
\M^{\mu\nu}(H) &= i\frac{g^2 (p_2\cdot V) g^{\mu\nu}}{4(s-m_H^2)}
\crr\cr
\M^{\mu\nu}(\nu^\ast) &= \frac{-ie^2p_3^\alpha
p_4^\beta}{2\Lambda^2(t-m_{\nu^\ast}^2)}
\Big( (c^2-d^2)\trace{\s^{\nu\beta}
(\pslash_3-\pslash_1) \s^{\mu\alpha}(\Vslash + \Vslash^5
\gamma_5)} -\cr
&\qquad\qquad
\frac{m_{\nu^\ast}}{m_e}\trace{(
c^2+d^2-2cd\gamma_5) \s^{\nu\beta}\s^{\mu\alpha}(\Vslash + \Vslash^5
\gamma_5) \pslash_2} \Big).
}\eqn\MMsII
$$
Equation \Mll\ together with equations \GammaV-\MMsII\ allow a
straight-forward, if lengthy, calculation of the various helicity
amplitudes.

\chapter{Conclusion}

To summarize, we have developed a technique for calculating Feynman
amplitudes involving (possibly massive) fermions. The technique
uses two (complex) four-vectors $V$ and $V_5$ which depend
on the fermion momenta and helicities. Equations~\OddFinal\ and
\EvenFinal\ contain the prescription for expressing any Feynman
amplitude as a trace involving these two four-vectors.

In addition to the calculation of tree-level amplitudes with massless
or massive fermions, the method can also be used in the calculation
quantities arising in loop calculations provided the spinors can be
taken to be in 4 dimensions.

The Vector Equivalence technique easily lends itself to computerized
evaluation of helicity amplitudes. The HIP package\refmark{\HIP}
implements the method symbolically.

\appendix

Expressing the four-vectors $V$ and $V_5$ in terms of the fermion
momenta can only be done in the context of a specific spinor
representation. For completeness we provide a full description of one
such representation. Our description closely follows that of reference
\Hagiwara.

The gamma matrices are given by
$$
\gamma^\mu = \pmatrix{0 & \s^\mu_+ \cr \s^\mu_- & 0 \cr},\qquad
\gamma^5 = \pmatrix{-1 & 0 \cr 0 & 1\cr},\eqn\ExplicitGammas
$$
where $\s^\mu_\pm = (1, \pm\bs)$, and
$$
\bs = (\s^1, \s^2, \s^3) = \left[\pmatrix{0 & 1 \cr 1 & 0\cr},
\pmatrix{0 & -i \cr i & 0 \cr}, \pmatrix{1 & 0 \cr 0 &
-1\cr}\right].\eqn\ExplicitSigmas
$$

The spinors $u(p, \la)$ and $v(p, \la)$ are given by
$$\eqalign{
u(p, \la) &= \pmatrix{u(p, \la)_- \cr u(p, \la)_+ \cr}, \qquad \bar
u(p, \la) = \pmatrix{u(p, \la)_+^\dagger & u(p, \la)_-^\dagger \cr},
\cr
v(p, \la) &= \pmatrix{v(p, \la)_- \cr v(p, \la)_+\cr}, \qquad \bar
v(p, \la) = \pmatrix{v(p, \la)_+^\dagger & v(p, \la)_-^\dagger \cr}.
}\eqn\ExplicitUV
$$
The explicit components of $u_\pm$ and $v_\pm$ are given by
$$\eqalign{
u(p, \la)_\pm = \omega_\pm(p) \chi_\la(p). \\
v(p, \la)_\pm = \pm\la\omega_\la(p) \chi_{-\la}(p),
}\eqn\UPlusMinus
$$
where $\omega_\pm(p) = \sqrt{E\pm |\bp|}$ and $\chi_\pm(p)$
are the helicity eigenstates
$$
\frac{\bs\cdot \bp}{|\bp|} \chi_\la(p) = \la
\chi_\la(p),\eqn\ChiEigen
$$
and are given by
$$\eqalign{
\chi_+(p) &= \frac{1}{\sqrt{2|\bp|( |\bp| + p_z )}}
\pmatrix{ |\bp| + p_z \cr p_x + ip_y\cr} = \frac{1}{\sqrt{2}}\pmatrix{
\sqrt{1+\cos\th} \cr \sqrt{1-\cos\th} e^{i\phi}\cr}
\cr
\chi_-(p) &= \frac{1}{\sqrt{2|\bp|( |\bp| + p_z )}}
\pmatrix{ -p_x + ip_y \cr |\bp| + p_z \cr} =
\frac{1}{\sqrt{2}}\pmatrix{-\sqrt{1-\cos\th}e^{-i\phi} \cr
\sqrt{1+\cos\th}\cr}
}\eqn\ExplicitChi
$$
For an arbitrary momentum $p^\mu = (E, \bp)$ where
$$
\bp = (p_x, p_y, p_z) = \left(\sin\th\cos\phi |\bp|,\,\,
\sin\th\sin\phi |\bp|,\,\, \cos\th |\bp|\right).\eqn\ExplicitP
$$
In the special case of $\th=\pi$ ($p_z = -|\bp|$) we use
$$
\chi_+(p) = \pmatrix{ 0 \cr 1\cr},\qquad
\chi_-(p) = \pmatrix{ -1 \cr 0 \cr}.\eqn\SpecialChi
$$

The equations in this appendix, together with eqn.~\VviaG\ can be used
to calculation $V$, $V_5$ and $V_\la$ in terms of the fermion momenta
and helicities. The result of the calculation is:
\def\pprime{{}^{\scriptscriptstyle(}{}'{}^{\scriptscriptstyle)}}
$$
\eqalign{
V &= {1\over4}\left(|{\bp}|\,|{\bp}'|\,p_0\,p'_0\right)^{-1/2}\tilde
V,\cr
V^5 &= {s\over4}\left(|{\bp}|\,|{\bp}'|\,p_0\,p'_0\right)^{-1/2}\tilde V^5}
\eqn\Final
$$

$$
\left.
\eqalign{
\tilde V_x &= (-p_-p'_-+p_+p'_+)(p'_0p_\bot +p_0p'_\bot)\cr
\tilde V_y &= is (-p_-p'_-+p_+p'_+)(p'_0p_\bot -p_0p'_\bot)\cr
\tilde V_z &= (p_-p'_--p_+p'_+)(p_\bot p'_\bot-p_0p'_0)\cr
\tilde V_t &= (p_-p'_-+p_+p'_+)(p_\bot p'_\bot+p_0p'_0)
}
\right\}\quad (s=s'),
\eqn\FinalI
$$

$$
\left.
\eqalign{
\tilde V_x &= (p_-p'_++p_+p'_-)(-p_\bot p'_\bot+p_0p'_0)\cr
\tilde V_y &= -is(p_-p'_++p_+p'_-)(p_\bot p'_\bot+p_0p'_0) \cr
\tilde V_z &= -(p_-p'_++p_+p'_-)(p'_0p_\bot +p_0p'_\bot)\cr
\tilde V_t &= (-p_-p'_++p_+p'_-)(p'_0p_\bot -p_0p'_\bot)
}
\right\}\quad (s=-s'),
\eqn\FinalII
$$

$$
\left.
\eqalign{
\tilde V^5_x &= (p_-p'_-+p_+p'_+)(p'_0p_\bot +p_0p'_\bot) \cr
\tilde V^5_y &= is (p_-p'_-+p_+p'_+)(p'_0p_\bot  - p_0p'_\bot) \cr
\tilde V^5_z &= -(p_-p'_-+p_+p'_+)(p_\bot p'_\bot -p_0p'_0)\cr
\tilde V^5_t &= -(p_-p'_--p_+p'_+)(p_\bot p'_\bot + p_0p'_0)
}
\right\}\quad (s=s'),
\eqn\FinalIII
$$

$$
\left.
\eqalign{
\tilde V^5_x &= (p_-p'_+-p_+p'_-)(p_\bot p'_\bot - p_0p'_0) \cr
\tilde V^5_y &= -is (-p_-p'_++p_+p'_-)(p_\bot p'_\bot + p_0p'_0) \cr
\tilde V^5_z &= (p_-p'_+-p_+p'_-)(p'_0p_\bot +p_0p'_\bot) \cr
\tilde V^5_t &= (p_-p'_++p_+p'_-)(p'_0p_\bot -p_0p'_\bot)
}
\right\}\quad (s=-s'),
\eqn\FinalIV
$$
where $s\pprime$ is the helicity of $p\pprime$ ($-{\rm helicity}$ for
an anti-fermion), $c\pprime$ is 1 (-1) for an (anti) fermion,
$$\eqalign{
|\bp\pprime| &= \left[{p_x\pprime}^2 +{p_y\pprime}^2
+{p_z\pprime}^2\right]^{1/2},\cr
{p_0\pprime} &= |{\bf p}\pprime|+p_z\pprime,\cr
p_-\pprime &= \left[E\pprime -|{\bf p}\pprime|\right]^{1/2},\cr
p_+\pprime &= s\pprime c\pprime\left[E\pprime +|{\bf
p}\pprime|\right]^{1/2},\cr
p_\bot &= p_x- isp_y,\quad{\rm and}\qquad p'_\bot = p'_x+is'p'_y.
}
\eqn\FinalDefs
$$
In the limit $p_0\pprime, p_\bot\pprime\rightarrow0$, one should take
$p_\bot\pprime/\sqrt{p_0\pprime}\rightarrow\sqrt{2|\bp\pprime|}$.

\refout
\bigskip

\bye